\documentclass[aps,prd,reprint,amsmath,amssymb,longbibliography]{revtex4-2}

\usepackage[T1]{fontenc}
\usepackage[utf8]{inputenc}
\usepackage{lmodern}
\usepackage{microtype}
\usepackage{amsmath,amssymb,mathtools,bm}
\usepackage{booktabs,tabularx,array}
\usepackage{xcolor}
\usepackage{hyperref}
\usepackage{xurl}

\hypersetup{
  colorlinks=true,
  linkcolor=blue!45!black,
  citecolor=green!35!black,
  urlcolor=blue!45!black,
  pdftitle={Characteristic-cone running from on-shell residues: Identifiability on the physical EFT quotient},
  pdfauthor={Renat Gafarov},
  pdfsubject={Physical EFT quotients, characteristic response, and amplitude-data completeness},
  pdfkeywords={effective field theory, characteristics, on-shell amplitudes, Schur complement, identifiability}
}

\newcolumntype{Y}{>{\raggedright\arraybackslash}X}
\newcommand{\Mp}{M_{\mathrm P}}
\newcommand{\bmu}{\beta_{\mu}}

\newcommand{\Xb}{\mathcal X}
\newcommand{\CW}{C_{W^2}}
\newcommand{\CQ}{C_{\gamma\phi}^{(2)}}
\newcommand{\dd}{\mathrm d}
\newcommand{\MSbar}{\overline{\mathrm{MS}}}
\newcommand{\order}{\mathcal O}
\newcommand{\Efour}{\mathfrak E_4}

\newcommand{\Wflow}{\mathfrak W_{\rm adm}}
\newcommand{\Wbeta}{\mathfrak W_{\beta}}
\newcommand{\Ephys}{E_{\mathcal B}}
\newcommand{\Amap}{\mathcal A_{J=2}}
\newcommand{\Xmap}{\boldsymbol\chi_{\mathcal B}}
\newcommand{\lambdaC}{\lambda_{\mathcal C}}

\begin{document}

\title{Characteristic-cone running from on-shell residues: Identifiability on the physical EFT quotient}
\author{Renat Gafarov}
\email{renatusb79@gmail.com}
\affiliation{Independent Researcher, Cetinje, Montenegro}
\date{1 August 2026}

\newcommand{\manuscriptabstract}{%
We formulate a data-completeness test for extracting low-energy
characteristic-cone running from flat-space on-shell renormalization-group
residues.  In four-dimensional, parity-even Einstein--scalar--Maxwell
effective field theory through four derivatives and four external fields,
Wilson directions are first placed on the physical EFT quotient and
background propagation is represented by a traceless response on the full
massless root sector.  The response is identifiable from chosen amplitude
coordinates precisely when the amplitude kernel lies in the response
kernel.  For the published opposite-helicity vector--scalar residue we find
$\beta_\mu C_{\gamma\phi}^{(2)}=-55/(96\pi^2)$ and, after an order-reduced
frame translation, the contribution
$\beta_\mu(r_T-r_\gamma)=-55/(48\pi^2M_{\rm P}^4)$.  A finite operator
registry gives explicit rank-two incompleteness tests: independent $F^4$
directions split photon polarizations on a magnetic background, while
$\mathcal CFF$ produces an analytic splitting along a principal Weyl axis
on a Petrov-I Einstein--scalar background.  For the isolated
scalar--photon contact, direct Schur reduction finds nonzero raw mixing but
no first-order physical off-diagonal entry, by the Maxwell Ward identity.
Thus the residue fixes one diagonal cone contribution but does not determine
the term-complete characteristic response over the declared admissible
coefficient domain.
}
\begin{abstract}\manuscriptabstract\end{abstract}
\maketitle

\section{Introduction}
\label{sec:introduction}

Low-energy propagation cones in an effective field theory (EFT) are physical
data only after gauge directions, constraints, redundant operators, and
common frame shifts have been removed.  Their renormalization-group (RG) flow
therefore cannot in general be read from a single component of an off-shell
counterterm basis.  On-shell methods bypass much of that bookkeeping, but a
different question then arises: how much background characteristic
information is identifiable from a specified set of on-shell residues?

The setting is motivated by post-GW170817 scalar--tensor EFTs whose
tree-level tensor cone is luminal, conventionally represented in the
Horndeski basis by $G_{4,X}=0$ and $G_5=0$
\cite{LIGOScientific:2017zic,Creminelli:2017sry}.  Loop effects can separate
low-energy light and gravity cones \cite{deRham:2019ctd,deRham:2020zyh}.
Here the input is the opposite-helicity vector--scalar $J=2$ anomalous
dimension of Ref.~\cite{Baratella:2021lwh}.  We specialize it to one real
scalar and one Abelian vector, translate it to an order-reduced covariant
frame, and determine its contribution to the relative tensor--photon cone.

The data question is sharper than operator counting.  We define the
amplitude-coordinate map and the background characteristic-response map on
the same physical EFT quotient.  The familiar factorization condition
$X=\widetilde X\circ A$ then becomes an operational completeness test:
$\ker A\subseteq\ker X$.  A finite registry makes the test explicit.  On a
magnetic background, the selected amplitude row and a photon-splitting row
have stacked rank two because of independent $F^4$ directions.  On a
Weyl-curved background, an analytic $\mathcal CFF$ response supplies a
second rank-two test.  These statements concern inference over a declared
coefficient domain; they do not assert that every witness component occurs
in the beta vector of every fixed UV completion.

Section~\ref{sec:eft} declares the EFT domain and local-solution backgrounds.
Section~\ref{sec:method} constructs the quotient-level response and proves
its field-coordinate invariance.  Section~\ref{sec:onshell} gives the
published-residue normalization and frame map.  Section~\ref{sec:em-test}
performs the single-contact Schur/Ward calculation.
Sections~\ref{sec:witnesses} and \ref{sec:petrov} give the finite rank tests
and the explicit Petrov-I realization, followed by the scope and conclusions.

\section{EFT, backgrounds, and scheme}
\label{sec:eft}

We use signature $(-,+,+,+)$ and
\begin{equation}
 S_0=\int\!\dd^4x\sqrt{-g}\left[
 \frac{\Mp^2}{2}R-\frac12\nabla_\mu\phi\nabla^\mu\phi
 -\frac14F_{\mu\nu}F^{\mu\nu}\right],
 \label{eq:leading-action}
\end{equation}
where $\Mp$ is the reduced Planck mass.  Write
$V_\mu=\nabla_\mu\phi$, $\bar V_\mu=\nabla_\mu\bar\phi$, and
\begin{equation}
 X_H=-\frac12V^2,\qquad \Xb=-\bar V^2>0,
 \qquad \bmu=\mu\frac{\dd}{\dd\mu}.
 \label{eq:conventions}
\end{equation}

The declared domain is four dimensional, local, parity even, and invariant
under diffeomorphisms, the Abelian gauge symmetry, and the scalar shift
$\phi\mapsto\phi+\text{const}$.  It contains one real massless scalar and one
Abelian vector, is truncated at four derivatives and at four external fields,
and uses one-loop massless $\MSbar$ amplitude coordinates.  Integrations by
parts, topological densities, leading equations of motion, and perturbatively
invertible field redefinitions are quotiented.  The finite directions needed
below are listed in Table~\ref{tab:registry}; the registry is a test sector,
not a claim that these four entries exhaust every four-derivative observable.

\begin{table}[t]
\caption{Normalized directions in the finite data-completeness test.  A zero
in the last column means that the selected opposite-helicity
$\gamma^-\gamma^+\phi\phi$ coordinate does not see that direction.}
\label{tab:registry}
\begin{tabularx}{\columnwidth}{@{}lYYc@{}}
\toprule
Direction & Quotient class & Independent on-shell class & $\Amap$ \\
\midrule
$\bm w_{\gamma\phi}$ &
$V_\mu V_\nu T_\gamma^{\mu\nu}/\Mp^4$ &
$\gamma^-\gamma^+\phi\phi$ & $1$ \\
$\bm w_a$ & $(F_{\mu\nu}F^{\mu\nu})^2/\Lambda_F^4$ &
four photon & $0$ \\
$\bm w_b$ & $(F_{\mu\nu}\widetilde F^{\mu\nu})^2/\Lambda_F^4$ &
four photon & $0$ \\
$\bm w_{\mathcal C}$ &
$\lambdaC\mathcal C_{\mu\nu\rho\sigma}F^{\mu\nu}F^{\rho\sigma}$ &
same-helicity $h\gamma\gamma$ & $0$ \\
\bottomrule
\end{tabularx}
\end{table}

For a fixed EFT the beta function at a point is one tangent vector, not a
vector space.  We denote by $\Wflow$ the linear admissible coefficient tangent
domain allowed by the declared assumptions, and by
$\Wbeta\subseteq\Wflow$ the linear span of beta vectors realized by a
specified theory family or matching class.  The question studied here is
uniform inference over $\Wflow$.  Failure there means that the selected datum
is insufficient without further model information; it does not establish
that a particular fixed beta vector has a nonzero component along every
witness direction.

All backgrounds are frozen local jets of solutions of the leading equations
from Eq.~\eqref{eq:leading-action}.  In Riemann normal coordinates we impose
the WKB hierarchy
\begin{equation}
 |R|,\quad \frac{|\nabla\bar V|^2}{|\bar V|^2}
 \ll |q|^2\ll\Lambda_{\rm EFT}^2,
 \label{eq:wkb}
\end{equation}
together with perturbatively small Wilson corrections.  Thus a locally
constant $\bar V$ or $\bar F$ is a frozen-coefficient approximation to a
weakly curved leading-order solution, not a global flat solution with
nonzero stress tensor.  The Petrov-I example in Sec.~\ref{sec:petrov} is an
exact leading Einstein--scalar solution.

The logarithmic input is a renormalized four-dimensional external-state
on-shell coordinate obtained after dimensional regularization and massless
$\MSbar$ subtraction.  Four-dimensional Gauss--Bonnet identities and order
reduction below are applied to that renormalized physical class, not to an
arbitrary $d=4-2\epsilon$ off-shell coefficient.  Evanescent operators can
alter off-shell poles and higher-loop finite information, but no such
higher-loop identity is asserted here \cite{Bern:2015xsa}.

\section{Physical characteristic maps and identifiability}
\label{sec:method}

\subsection{Generalized pencils and Schur reduction}

On a background with one preferred direction, a gauge-reduced principal
symbol may take the form
\begin{align}
 Q&=g^{\mu\nu}q_\mu q_\nu,
 &W&=(\bar V^\mu q_\mu)^2,
 \nonumber\\
 K(q)&=K_QQ+K_WW.
 \label{eq:pencil}
\end{align}
The cone parameters are generalized eigenvalues,
\begin{equation}
 K_Wv_A=r_AK_Qv_A,\qquad
 l_A^{\mathsf T}K_W=r_A l_A^{\mathsf T}K_Q.
 \label{eq:gen-eigen}
\end{equation}
For a simple branch, differentiating the pencil gives
\begin{equation}
 \boxed{
 \bmu r_A=
 \frac{l_A^{\mathsf T}(\bmu K_W-r_A\bmu K_Q)v_A}
 {l_A^{\mathsf T}K_Qv_A}.}
 \label{eq:beta-r}
\end{equation}
At a $d$-fold semisimple root, choose dual bases with
$L_r^{\mathsf T}K_QV_r=\mathbf1_d$.  The first-order slopes are the
eigenvalues of
\begin{equation}
 \mathsf H_r=L_r^{\mathsf T}(\bmu K_W-r\bmu K_Q)V_r,
 \label{eq:degenerate-H}
\end{equation}
not its diagonal entries in an arbitrary basis.

Gauge directions are first quotiented, or fixed on a regular chart;
nondynamical constraint variables are then eliminated by a Schur complement.
For a split of the gauge-invariant action symbol, work where the auxiliary
block $D$ is invertible and away from roots of $D$:
\begin{equation}
 P_*=
 \begin{pmatrix}A&B\\ C&D\end{pmatrix},\qquad
 K_{\rm phys}=A-BD^{-1}C,
 \label{eq:schur}
\end{equation}
the flow is
\begin{align}
 \bmu K_{\rm phys}={}&\bmu A-(\bmu B)D^{-1}C
 -BD^{-1}(\bmu C)
 \nonumber\\
 &+BD^{-1}(\bmu D)D^{-1}C.
 \label{eq:beta-schur}
\end{align}
On a general anisotropic background, define a root at fixed direction by
$P_{\rm phys}(c,\hat{\bm q};\mu)v_A=0$.  A simple root obeys
\begin{equation}
 \boxed{
 \bmu c_A=-
 \frac{l_A^{\mathsf T}(\bmu P_{\rm phys})v_A}
 {l_A^{\mathsf T}(\partial_cP_{\rm phys})v_A}.}
 \label{eq:root-flow}
\end{equation}

\subsection{The physical quotient and the identifiability criterion}

Let $\Efour$ be the four-derivative EFT quotient declared in
Sec.~\ref{sec:eft}.  Correlated terms generated by a redefinition are
retained: keeping only one convenient remainder would not define an element
of the quotient.  The admissible tangent domain is
$\Wflow\subseteq T_{[C]}\Efour$.  Higher-derivative representatives are
order reduced before their light-branch characteristics are evaluated.  If
additional symmetry, matching, or UV information restricts the problem to
$\Wbeta$, the criterion must be reapplied on that smaller domain.

At a fixed background $\mathcal B$ and wave-vector direction, suppose the
leading physical symbol has a $d$-fold semisimple root $c_0$.  Define its
selected physical root sector by
\begin{equation}
 \Ephys=\ker P_{\rm phys}(c_0,\hat{\bm q}),
 \label{eq:physical-root-space}
\end{equation}
and choose right and left null bases $V,L$ after the gauge quotient and Schur
reduction.  In the Einstein--scalar--Maxwell examples the unperturbed common
root is always the full five-dimensional physical sector
\begin{equation}
 \Ephys=\operatorname{span}(h_+,h_\times,\varphi,A_2,A_3),
 \label{eq:five-sector}
\end{equation}
for propagation along the first tetrad axis; photon-only quantities below are
linear projections of the response on this full sector.  Define
\begin{align}
 \mathsf D_{\mathcal B}
 &=L^{\mathsf T}(\partial_cP_{\rm phys})V,
 \nonumber\\
 \mathsf N_{\mathcal B}(\bm b)
 &=L^{\mathsf T}(\delta_{\bm b}P_{\rm phys})V,
 \nonumber\\
 \mathsf R_{\mathcal B}(\bm b)
 &=-\mathsf D_{\mathcal B}^{-1}\mathsf N_{\mathcal B}(\bm b),
 \label{eq:root-endomorphism}
\end{align}
assuming $\mathsf D_{\mathcal B}$ is invertible.  The eigenvalues of
$\mathsf R_{\mathcal B}$ are the first-order root slopes.  Under a change of
null bases, this endomorphism transforms by similarity.

A common displacement of every branch is removed before defining the linear
physical characteristic map,
\begin{equation}
 \boxed{
 \Xmap(\bm b)=\mathsf R_{\mathcal B}(\bm b)
 -\frac1d\operatorname{tr}\mathsf R_{\mathcal B}(\bm b)\,\mathbf1_d.}
 \label{eq:physical-characteristic-map}
\end{equation}
Thus $\Xmap:\Wflow\to\operatorname{End}_0(\Ephys)$ is linear and is defined
up to conjugation.  Its kernel is intrinsic.  It retains not only relative
eigenvalue slopes but also a possible traceless nilpotent perturbation that
could affect semisimplicity without shifting eigenvalues at first order.

The map descends to the perturbative EFT quotient.  Under a local,
perturbatively invertible field redefinition, with the leading background
solution transformed at the same time, the first-order quadratic symbol has
the form
\begin{equation}
 \delta P\longmapsto\delta P+\mathsf A P_0+P_0\mathsf B
 \label{eq:symbol-redefinition}
\end{equation}
up to the leading background equations and higher EFT order.  Since
$P_0V=0$ and $L^{\mathsf T}P_0=0$, both added terms vanish in
$L^{\mathsf T}\delta P V$.  Order reduction retains this light branch and
discards the spurious heavy roots of an unreduced derivative representative.
Accordingly, Eq.~\eqref{eq:physical-characteristic-map} is invariant to the
working order.  It is a low-energy, order-reduced characteristic, not the
characteristic polynomial of the unreduced higher-derivative PDE
\cite{Reall:2021voz,Figueras:2024dta,Gavassino:2026abc,Thaalba:2026abc}.

The selected on-shell coordinate is the linear map
\begin{equation}
 \Amap:\Wflow\longrightarrow\mathbb R,
 \qquad \Amap(\bm b)=b_{\gamma\phi}
 \equiv(\bmu\CQ)_{\bm b}.
 \label{eq:amplitude-map}
\end{equation}

\paragraph*{Identifiability criterion.}
The physical characteristic response $\Xmap(\bm b)$ is uniquely determined
by the single datum $\Amap(\bm b)$ for every $\bm b\in\Wflow$ if and only if
\begin{equation}
 \boxed{\ker\Amap\subseteq\ker\Xmap.}
 \label{eq:identifiability}
\end{equation}
Equivalently, there is a unique linear map
$\overline\chi_{\mathcal B}:\operatorname{im}\Amap\to
\operatorname{End}_0(\Ephys)$ such that
$\Xmap=\overline\chi_{\mathcal B}\circ\Amap$.  Since both maps are defined
on the EFT quotient and the codomain contains only a physical relative
endomorphism, this condition is unchanged by invertible changes of operator
or physical-root coordinates.  Appendix~\ref{app:theorem} gives the proof.

For a finite operator basis, let $\mathsf A$ be the matrix of measured
amplitude coordinates and let $\mathsf X$ be any matrix representation of
the characteristic response.  The condition is the directly checkable rank
identity
\begin{equation}
 \boxed{\operatorname{rank}\mathsf A=
 \operatorname{rank}\!\begin{pmatrix}\mathsf A\\\mathsf X\end{pmatrix}.}
 \label{eq:rank-test}
\end{equation}
Failure for even one physical linear response functional proves failure for
the full response.  The two witnesses below use polarization splitting as
that functional.

The equivalence in Eq.~\eqref{eq:identifiability} is the standard
factorization lemma for linear maps.  Its role here is diagnostic rather than
mathematically novel: the nontrivial physics lies in defining $\Xmap$ on the
order-reduced, gauge-reduced quotient and in specifying the admissible domain
$\Wflow$.

Failure of Eq.~\eqref{eq:identifiability} does not say that a physical beta
function does not exist.  It says that the beta function is not recoverable
from the specified input.  Additional amplitude coordinates can restore
identifiability by replacing $\Amap$ with a larger map whose kernel is
smaller.

\section{On-shell input and order-reduced frame map}
\label{sec:onshell}

The amplitude coordinate of Ref.~\cite{Baratella:2021lwh} is
\begin{equation}
 \mathcal A(1V^-,2V^+,3\phi,4\bar\phi)
 =-\frac{\CQ}{\Mp^4}\langle13\rangle^2[23]^2.
 \label{eq:amplitude}
\end{equation}
Its Eqs.~(73)--(74) give
\begin{equation}
 \bmu\CQ=-\frac1{8\pi^2}
 \left(K+\frac{131}{30}\right),\qquad
 K=\frac{N_\phi}{30}+\frac{N_\psi}{20}+\frac{N_V}{5}.
 \label{eq:baratella}
\end{equation}
Here $N_\phi$ counts complex scalars.  One real scalar therefore contributes
$N_\phi=1/2$, while one Abelian vector contributes $N_V=1$.  Hence
\begin{equation}
 K=\frac1{60}+\frac15=\frac{13}{60},\qquad
 \boxed{\bmu\CQ=-\frac{55}{96\pi^2}}.
 \label{eq:beta-C}
\end{equation}
The replacement $N_\phi=1/2$ concerns loop multiplicity.  The coefficient in
Eq.~\eqref{eq:amplitude} is fixed by the two-real-scalar matrix element of the
contact below, so no additional LSZ factor is inserted.

To make the convention translation explicit, consider
\begin{multline}
 S=S_0+\int\!\dd^4x\sqrt{-g}\bigg[
 \CW W_{\mu\nu\rho\sigma}W^{\mu\nu\rho\sigma}
 \\
 {}+\frac{C_4}{\Mp^2}G^{\mu\nu}V_\mu V_\nu\bigg].
 \label{eq:frame-action}
\end{multline}
At first order in the Wilson coefficients, the metric redefinitions
\begin{align}
 \delta_W g_{\mu\nu}&=\frac{4\CW}{\Mp^2}
 \left(R_{\mu\nu}-\frac16Rg_{\mu\nu}\right),
 \nonumber\\
 \delta_4 g_{\mu\nu}&=\frac{2C_4}{\Mp^4}V_\mu V_\nu
 \label{eq:metric-redefs}
\end{align}
remove the non-topological $W^2$ part and the John operator, respectively,
up to leading equations of motion and higher EFT order.  With
$\delta S_m=\frac12\int\sqrt{-g}\,T^{\mu\nu}\delta g_{\mu\nu}$, they induce
\begin{align}
 \Delta\mathcal L_{W^2}&=\frac{2\CW}{\Mp^4}
 \left(T_{\mu\nu}T^{\mu\nu}-\frac13T^2\right),
 \nonumber\\
 \Delta\mathcal L_4&=\frac{C_4}{\Mp^4}
 V_\mu V_\nu T^{\mu\nu}.
 \label{eq:induced-contacts}
\end{align}
The Maxwell stress tensor is traceless and
$T^\phi_{\mu\nu}T_\gamma^{\mu\nu}=V_\mu V_\nu
T_\gamma^{\mu\nu}$.  The scalar--photon contact is therefore
\begin{align}
 \Delta\mathcal L_{\phi\gamma}^{J=2}
 &=\frac{C_4+4\CW}{\Mp^4}
 V_\mu V_\nu T_\gamma^{\mu\nu},
 \nonumber\\
 &\boxed{\CQ=C_4+4\CW.}
 \label{eq:frame-map}
\end{align}
The same $W^2$ redefinition also generates four-photon and four-scalar
contacts through Eq.~\eqref{eq:induced-contacts}.  In the on-shell coordinate
basis these are independent $4\gamma$ and $4\phi$ coordinates.  Therefore a
derivative with respect to $\CQ$ means varying the
$2\gamma2\phi$ coordinate while holding the independent $4\gamma$ and
$4\phi$ coordinates fixed.  Equation~\eqref{eq:frame-map} is precisely the
$2\gamma2\phi$ projection of the correlated off-shell redefinition, not a
license to discard the other projections.

The same physical combination appears in the speed and amplitude analysis of
Ref.~\cite{deRham:2019ctd}; Eq.~\eqref{eq:frame-map} is a cross-convention
rederivation matched to the $J=2$ normalization of
Ref.~\cite{Baratella:2021lwh}, not a new operator identity.  It is an equality
in the order-reduced on-shell class, not a componentwise identity between
arbitrary off-shell bases.  Extra roots generated by a derivative
representative lie outside the EFT branch and are discarded in the same
perturbative order reduction used to define $\Efour$.

In the tensor frame the contact gives
\begin{equation}
 G_\gamma^{\mu\nu}\doteq g^{\mu\nu}
 -\frac{2C}{\Mp^4}\bar V^\mu\bar V^\nu,\qquad C\equiv C_4+4\CW,
 \label{eq:photon-metric}
\end{equation}
so $r_\gamma=-2C/\Mp^4$ and $r_T=0$ in that frame.  A common running shift
$r_A\mapsto r_A+s(\mu)$ changes absolute beta functions but not their
difference.

On the baseline common cone, the physical matrix is diagonal in the full
sector $(h_+,h_\times,\varphi,A_2,A_3)$.  The selected residue therefore
contributes
\begin{equation}
 \boxed{
 \left.\bmu(r_T-r_\gamma)\right|_{J=2\;\mathrm{residue}}
 =-\frac{55}{48\pi^2\Mp^4}.}
 \label{eq:beta-delta}
\end{equation}
The qualifier is essential: Eq.~\eqref{eq:beta-delta} is not asserted to be
the term-complete $p^4$ background beta function.

The normalization chain is
\begin{align}
 \widehat\Delta&=\Mp^4(r_T-r_\gamma)=2C,
 \nonumber\\
 \bmu\widehat\Delta&=-\frac{55}{48\pi^2},
 \nonumber\\
 \left.\bmu(c_T^2-c_\gamma^2)\right|_{\Delta=0}
 &=-\frac{55}{48\pi^2}\frac{\Xb}{\Mp^4},
 \nonumber\\
 \left.\bmu(c_T-c_\gamma)\right|_{\Delta=0}
 &=-\frac{55}{96\pi^2}\frac{\Xb}{\Mp^4}.
 \label{eq:normalization-chain}
\end{align}

\section{Restricted single-contact electromagnetic Schur analysis}
\label{sec:em-test}

At two derivatives, background-induced $h$--$A$ and $h$--$\phi$ terms in
minimal Einstein--scalar--Maxwell theory carry one fluctuation derivative and
are subprincipal relative to the $q^2$ diagonal blocks.  Thus
$\bar F_{\mu\nu}\neq0$ alone does not ensure principal mixing.

The contact in Eq.~\eqref{eq:frame-map} does have a principal quadratic
Hessian when both $\bar V$ and $\bar F$ are nonzero.  With
$u_\mu=q_\mu\varphi$, $f_{\mu\nu}=q_\mu a_\nu-q_\nu a_\mu$, and
\begin{equation}
 \tau^{\mu\nu}(\bar F,f)=
 \bar F^{\mu\rho}f^\nu{}_{\rho}
 +f^{\mu\rho}\bar F^\nu{}_{\rho}
 -\frac12g^{\mu\nu}\bar F_{\rho\sigma}f^{\rho\sigma},
\end{equation}
its principal quadratic part is
\begin{align}
 \delta^2(V_\mu V_\nu T_\gamma^{\mu\nu})\doteq{}&
 u_\mu u_\nu\bar T_\gamma^{\mu\nu}
 +2\bar V_\mu u_\nu\tau^{\mu\nu}(\bar F,f)
 \nonumber\\
 &+\bar V_\mu\bar V_\nu T_\gamma^{\mu\nu}(f).
 \label{eq:background-hessian}
\end{align}

Choose a local tetrad
\begin{align}
 q_\mu&=(-\omega,0,0,k),&
 \bar V_\mu&=(v,0,0,0),
 \nonumber\\
 \bar{\bm E}&=(E_\perp,0,E_\parallel),&
 \alpha&=C/\Mp^4,
\end{align}
and use $a_z=0$ as one regular coordinate chart on the Maxwell quotient.
Let $\lambda=\omega^2-k^2$ and $\Sigma=\omega^2+k^2$.  In the split
$u=(\varphi,a_x,a_y)$ and $z=a_0$, the single-contact principal Hessian is
\begin{align}
 A={}&\begin{pmatrix}
 \lambda+\alpha(E_\parallel^2\lambda+E_\perp^2\Sigma)&
 2\alpha E_\perp v\lambda&0\\
 2\alpha E_\perp v\lambda&\lambda+\alpha v^2\Sigma&0\\
 0&0&\lambda+\alpha v^2\Sigma
 \end{pmatrix},\nonumber\\
 B={}&\begin{pmatrix}2\alpha E_\parallel k v\omega\\0\\0\end{pmatrix},
 \qquad C_{\rm bl}=B^{\mathsf T},
 \nonumber\\[-2pt]
 D={}&k^2(1+\alpha v^2).
 \label{eq:explicit-blocks}
\end{align}
The label $C_{\rm bl}$ distinguishes the lower-left block from the Wilson
coefficient.  The Schur-reduced symbol is
\begin{equation}
 K_{\rm phys}=A-
 \frac{4\alpha^2E_\parallel^2v^2\omega^2}{1+\alpha v^2}
 \operatorname{diag}(1,0,0),
 \label{eq:explicit-schur}
\end{equation}
and therefore
\begin{equation}
 \left.\partial_\alpha(BD^{-1}C_{\rm bl})\right|_{\alpha=0}=0.
 \label{eq:schur-zero}
\end{equation}

This result is not tied to the gauge $a_z=0$.  The gauge-invariant Hessian
depends on $a_\mu$ only through $f_{\mu\nu}$ and hence annihilates the gauge
vector $a_\mu=q_\mu\chi$.  At the unperturbed Maxwell root $q^2=0$, it
induces a bilinear form on the two-dimensional physical photon space
\begin{equation}
 \mathcal V_\gamma(q)=
 \{a_\mu:q\!\cdot\!a=0\}/\{a_\mu\sim a_\mu+q_\mu\chi\}.
 \label{eq:photon-quotient}
\end{equation}
Regular gauge choices may change the representative matrix blocks, but they
induce the same bilinear form on $\mathcal V_\gamma(q)$.  Schur elimination in
different regular charts therefore represents that same induced form.
Equation~\eqref{eq:explicit-schur} is valid
for $k\neq0$ in the displayed chart, while the quotient statement supplies
the regular continuation to directions for which that coordinate chart
degenerates.

On a free physical photon state, $q^2=0$ and $q\cdot a=0$.  Direct contraction
gives the bilinear Maxwell Ward identity
\begin{equation}
 q_\nu\tau^{\mu\nu}(\bar F,f)=0,
 \label{eq:stress-ward}
\end{equation}
for the locally constant background used in the principal-symbol extraction.
Consequently
\begin{equation}
 2\bar V_\mu q_\nu\tau^{\mu\nu}=0.
 \label{eq:projected-cross-zero}
\end{equation}
At $\alpha=0$, the normalized left and right vectors are the unit vectors in
$(\varphi,a_x,a_y)$, and
\begin{equation}
 \left.\partial_\alpha K_{\rm phys}\right|_{\lambda=0,\alpha=0}
 =2k^2\operatorname{diag}(E_\perp^2,v^2,v^2).
 \label{eq:em-projected-matrix}
\end{equation}
Thus, within this isolated contact,
\begin{equation}
 \bmu c_\phi^2=-\frac{2\bmu C}{\Mp^4}E_\perp^2,
 \qquad
 \bmu c_{\gamma,x}^2=\bmu c_{\gamma,y}^2
 =-\frac{2\bmu C}{\Mp^4}v^2.
 \label{eq:single-contact-slopes}
\end{equation}
The calculation has nonzero raw principal blocks but no first-order physical
off-diagonal entry and no first-order Schur contribution.  It is therefore a
restricted negative result, not a nontrivial demonstration of
Eq.~\eqref{eq:beta-r}.

\section{Explicit non-identifiability witnesses}
\label{sec:witnesses}

The kernel criterion can be tested without knowing every beta-function
coefficient.  It is enough to identify an independent EFT direction that is
invisible to $\Amap$ but changes a physical characteristic root.

\subsection{Electromagnetic background}

Consider nonlinear electrodynamics.  Background-field birefringence from
quartic electromagnetic invariants is standard
\cite{BialynickaBirula:1970,Adler:1971}; here it is used only as a kernel
witness:
\begin{align}
 \mathcal L_\gamma={}&-\frac14 I
 +\frac{a}{\Lambda_F^4}I^2
 +\frac{b}{\Lambda_F^4}J^2,
 \nonumber\\
 I={}&F_{\mu\nu}F^{\mu\nu},\qquad
 J=F_{\mu\nu}\widetilde F^{\mu\nu}.
 \label{eq:f4-action}
\end{align}
Let $\bm w_a,\bm w_b\in\Wflow$ denote the corresponding independent
operator-coordinate directions in the declared inference domain.  They
contain four photon fields and no local two-photon--two-scalar contact.
Therefore
\begin{equation}
 \Amap(\bm w_a)=\Amap(\bm w_b)=0.
 \label{eq:f4-amplitude-kernel}
\end{equation}
For example,
\begin{equation}
 \left.I^2\right|_{f^2}
 =2\bar I\,f_{\mu\nu}f^{\mu\nu}
 +4(\bar F_{\mu\nu}f^{\mu\nu})^2,
 \label{eq:f4-hessian}
\end{equation}
so the physical Hessian is polarization dependent.

More explicitly, take $\bar{\bm B}=B\hat{\bm z}$ and propagation along
$\hat{\bm x}$.  With $\epsilon_B=B^2/\Lambda_F^4$, the transverse potentials
$A_y,A_z$ have quadratic forms
\begin{align}
 \mathcal L_y^{(2)}={}&\frac12\bigl[
 (1-16a\epsilon_B)\omega^2
 -(1-48a\epsilon_B)k^2\bigr]A_y^2,
 \nonumber\\
 \mathcal L_z^{(2)}={}&\frac12\bigl[
 (1-16a\epsilon_B+32b\epsilon_B)\omega^2
 \nonumber\\[-2pt]
 &\hspace{23mm}{}-(1-16a\epsilon_B)k^2\bigr]A_z^2.
 \label{eq:f4-quadratic}
\end{align}
Thus
\begin{equation}
 c_y^2=1-32a\epsilon_B,\qquad
 c_z^2=1-32b\epsilon_B
 \label{eq:f4-speeds}
\end{equation}
to first order.  Using $c^2$ as a smooth local root coordinate at $c_0=1$
rescales first-order responses by two and leaves all kernels and ranks
unchanged.  On the full root sector in Eq.~\eqref{eq:five-sector}, use
the physical response functional
\begin{equation}
 x_B(\bm w)=\frac12\left[
 \delta_{\bm w}c_y^2-\delta_{\bm w}c_z^2\right].
 \label{eq:photon-splitting-functional}
\end{equation}
It is unchanged by a common shift of all five branches and is a linear
projection of $\Xmap$.  In the ordered basis
$(\bm w_{\gamma\phi},\bm w_a,\bm w_b)$,
\begin{equation}
 \mathsf A_{J=2}=\begin{pmatrix}1&0&0\end{pmatrix},\qquad
 \mathsf x_B=\begin{pmatrix}0&-16\epsilon_B&16\epsilon_B\end{pmatrix}.
 \label{eq:em-rank-rows}
\end{equation}
Hence, for $B\neq0$,
\begin{equation}
 \operatorname{rank}\mathsf A_{J=2}=1,
 \qquad
 \operatorname{rank}\!\begin{pmatrix}\mathsf A_{J=2}\\\mathsf x_B\end{pmatrix}=2.
 \label{eq:em-kernel-witness}
\end{equation}
This is a rank-tested physical birefringence witness, not merely a change of
raw symbol entries.  Four-photon divergences are available in the minimally
coupled inventory of Ref.~\cite{Baratella:2021lwh}; the present statement
does not require assigning them a value in a fixed model.

\subsection{Curved background}

Use
$R^\rho{}_{\sigma\mu\nu}=\partial_\mu\Gamma^\rho_{\nu\sigma}
-\partial_\nu\Gamma^\rho_{\mu\sigma}+\cdots$ and the projected
curvature--photon class
\begin{equation}
 \Delta\mathcal L_{\mathcal CFF}=\lambdaC
 \mathcal C_{\mu\nu\rho\sigma}F^{\mu\nu}F^{\rho\sigma}.
 \label{eq:cff-witness}
\end{equation}
It is an independent $h\gamma\gamma$ on-shell class.  In two-spinor notation
its helicity content is
\begin{equation}
 \mathcal C_{\alpha\beta\gamma\delta}
 F^{\alpha\beta}F^{\gamma\delta}
 +\bar{\mathcal C}_{\dot\alpha\dot\beta\dot\gamma\dot\delta}
 \bar F^{\dot\alpha\dot\beta}\bar F^{\dot\gamma\dot\delta},
 \label{eq:cff-helicity}
\end{equation}
so it creates same-helicity photon pairs.  The amplitude map $\Amap$ instead
measures the opposite-helicity $\gamma^-\gamma^+\phi\phi$ coordinate.  Thus
\begin{equation}
 \Amap(\bm w_{\mathcal C})=0.
 \label{eq:cff-amplitude-kernel}
\end{equation}
The Weyl projection is important here.  Treating
$R_{\mu\nu\rho\sigma}\widetilde F^{\mu\nu}\widetilde F^{\rho\sigma}$
as identical to the pure $\mathcal CFF$ direction would be incorrect: their
difference contains Ricci terms that generate correlated matter contacts
after order reduction with the leading Einstein equation.  The
vector--tensor Horndeski interaction \cite{Horndeski:1976gi} and the
Drummond--Hathrell effective action \cite{Drummond:1979pp} contain closely
related curvature--photon structures, but no such identification is used in
the kernel argument.  A coupled, order-reduced principal-symbol treatment of
the related vector--Horndeski interaction is given in
Ref.~\cite{Davies:2021frz}.

At $\bar F=0$, choose an orthonormal tetrad in which the Weyl tensor is purely
electric and
\begin{equation}
 \mathcal E_{ij}=\mathcal C_{\hat0 i\hat0 j}
 =\operatorname{diag}(\mathcal E_1,\mathcal E_2,\mathcal E_3),
 \qquad \sum_i\mathcal E_i=0.
 \label{eq:electric-weyl}
\end{equation}
The spatial components obey
$\mathcal C_{ijkl}=\delta_{ik}\mathcal E_{jl}
+\delta_{jl}\mathcal E_{ik}-\delta_{il}\mathcal E_{jk}
-\delta_{jk}\mathcal E_{il}$, and therefore
\begin{equation}
 \mathcal C_{\mu\nu\rho\sigma}F^{\mu\nu}F^{\rho\sigma}
 =4\left(\mathcal E_{ij}E_i^{\rm em}E_j^{\rm em}
 -\mathcal E_{ij}B_i^{\rm em}B_j^{\rm em}\right).
 \label{eq:cff-eb}
\end{equation}
Take $q_{\hat a}=(-\omega,k,0,0)$ and the physical gauge
$A_{\hat0}=A_{\hat1}=0$.  The quadratic photon Lagrangian is
\begin{align}
 \mathcal L_\gamma^{(2)}={}&\frac12\left[
 (1+8\lambdaC\mathcal E_2)\omega^2
 -(1+8\lambdaC\mathcal E_3)k^2\right]A_2^2
 \nonumber\\
 &+\frac12\left[
 (1+8\lambdaC\mathcal E_3)\omega^2
 -(1+8\lambdaC\mathcal E_2)k^2\right]A_3^2.
 \label{eq:cff-quadratic}
\end{align}
Hence
\begin{align}
 c_2^2&=1+8\lambdaC(\mathcal E_3-\mathcal E_2),
 &c_3^2&=1-8\lambdaC(\mathcal E_3-\mathcal E_2)
 \label{eq:cff-speeds}
\end{align}
to first order.  In the full basis
$(h_+,h_\times,\varphi,A_2,A_3)$ this unit coefficient direction gives
the response in the squared-speed root coordinate,
\begin{equation}
 \Xmap^{(c^2)}(\bm w_{\mathcal C})=
 8(\mathcal E_3-\mathcal E_2)
 \operatorname{diag}(0,0,0,1,-1).
 \label{eq:cff-symbol}
\end{equation}
The response is nonzero whenever the two transverse Weyl eigenvalues differ.
Equivalently, with
$x_{\mathcal C}=\tfrac12(\delta c_2^2-\delta c_3^2)$, the basis
$(\bm w_{\gamma\phi},\bm w_{\mathcal C})$ gives
\begin{equation}
 \begin{pmatrix}\mathsf A_{J=2}\\\mathsf x_{\mathcal C}\end{pmatrix}
 =\begin{pmatrix}1&0\\0&8(\mathcal E_3-\mathcal E_2)\end{pmatrix},
 \label{eq:cff-rank}
\end{equation}
which has rank two at a suitable Petrov-I point.

Equations~\eqref{eq:em-kernel-witness} and \eqref{eq:cff-symbol} turn the
previous dimension-counting concern into a physical non-identifiability
statement.  They do not determine the missing coefficients; they show that
the corresponding independent amplitude coordinates must be controlled to
identify the response over the declared domain.  The witnesses do not by
themselves prove that their beta components are nonzero in a particular
fixed model.

The conclusion is conditional on the declared domain $\Wflow$.  A symmetry
that removes one of these directions must be imposed before applying the
kernel test.  Conversely, allowing the independent four-photon and
$h\gamma\gamma$ amplitude classes while measuring only
Eq.~\eqref{eq:amplitude} leaves precisely the ambiguity exhibited above.

\section{Petrov-I analysis of the selected residue}
\label{sec:petrov}

Consider the Jacobs Einstein--massless-scalar background
\cite{Jacobs:1968}
\begin{equation}
 \dd s^2=-\dd t^2+\sum_{i=1}^3t^{2p_i}\dd x_i^2,
 \qquad \bar\phi=\phi_0+s\ln(t/t_0),
 \label{eq:jacobs}
\end{equation}
with
\begin{equation}
 \sum_i p_i=1,\qquad
 \sum_i p_i^2=1-\frac{s^2}{\Mp^2}.
 \label{eq:jacobs-relations}
\end{equation}
The magnetic Weyl tensor vanishes for this diagonal Bianchi-I geometry.  It
is Petrov I only where the three electric-Weyl eigenvalues are distinct;
axisymmetric and conformally flat limits are type D and O, respectively
\cite{Petrov:2000}.  We therefore do not call the background ``generic
Weyl.''

Writing $\delta=s^2/\Mp^2$, the eigenvalues in the convention of
Eq.~\eqref{eq:electric-weyl} are
\begin{equation}
 \mathcal E_i=\frac{p_i(1-p_i)-\delta/3}{t^2},
 \qquad
 \mathcal E_3-\mathcal E_2=\frac{p_1(p_3-p_2)}{t^2}.
 \label{eq:jacobs-weyl}
\end{equation}
For example,
$(p_1,p_2,p_3)=(0.6,0.3,0.1)$ and $\delta=0.54$ satisfy
Eq.~\eqref{eq:jacobs-relations} and give
\begin{equation}
 (\mathcal E_1,\mathcal E_2,\mathcal E_3)
 =\frac{(0.06,0.03,-0.09)}{t^2}.
 \label{eq:jacobs-example}
\end{equation}
Equation~\eqref{eq:cff-speeds} then becomes
\begin{equation}
 c_2^2=1-0.96\frac{\lambdaC}{t^2},\qquad
 c_3^2=1+0.96\frac{\lambdaC}{t^2},
 \label{eq:jacobs-splitting}
\end{equation}
an explicit curved-background kernel witness.

Writing $\bar\Phi_{\mu\nu}\equiv\nabla_\mu\nabla_\nu\bar\phi$, the
nonzero scalar spurions in an orthonormal frame are
\begin{equation}
 \bar V_{\hat0}=\frac{s}{t},\quad
 \bar\Phi_{\hat0\hat0}=-\frac{s}{t^2},\quad
 \bar\Phi_{\hat i\hat j}=-\frac{sp_i}{t^2}\delta_{ij},\quad
 \bar\Box\bar\phi=0.
 \label{eq:jacobs-spurions}
\end{equation}
The background equations correlate curvature and scalar-gradient spurions;
a single trajectory does not independently vary their scalar background
coefficients and therefore cannot separate them by background scaling alone.

For $\bar F=0$, the selected contact
$V_\mu V_\nu T_\gamma^{\mu\nu}$ changes only the photon block.  Its Hessian
contains $\bar V$ but neither the Weyl tensor nor
$\bar\Phi_{\mu\nu}$.  Hence replacing the baseline geometry by the Jacobs
background does not turn this one residue into a species-mixed pencil.  The
restricted statement is
\begin{quote}
\textbf{Restricted single-residue Petrov statement.}  If the only running four-derivative
coordinate retained is $C_{\gamma\phi}^{(2)}$ and $\bar F=0$, curvature
anisotropy does not create species-mixed principal flow from that coordinate.
\end{quote}
This statement is deliberately narrower than the full EFT.  The independent
$\mathcal CFF$ result in Eqs.~\eqref{eq:cff-rank} and
\eqref{eq:jacobs-splitting} shows why the complete Petrov response is not
identifiable from the selected residue.

\section{Relation to radiative-stability analyses}
\label{sec:literature}

De Rham and Tolley obtain, for a canonical scalar background,
\begin{equation}
 c_h^2-1=4(C_4+4\CW)\frac{-\dot H}{\Mp^2}
 +\order(H^4/\Mp^4).
 \label{eq:drt}
\end{equation}
With $-\dot H=\Xb/(2\Mp^2)$ this is the matter-frame form of
Eq.~\eqref{eq:frame-map}.  If $C(\Lambda)=0$, then
\begin{equation}
 C(\mu)=\frac{55}{96\pi^2}\ln\frac{\Lambda}{\mu}>0
 \qquad(\mu<\Lambda),
\end{equation}
so the infrared sign agrees with their forward-limit positivity result,
within assumptions that control the massless-graviton $t$-channel pole;
standard pole-subtracted positivity is not automatic in gravity
\cite{Alberte:2020jsk}.  The general connection between analyticity,
positivity, and background superluminality goes back to
Ref.~\cite{Adams:2006sv}.\footnote{The
text immediately following Eq.~(4.50) of Ref.~\cite{deRham:2019ctd} writes
$C_4=-8C_{W^2}$, whereas that displayed equation requires
$C_4=-4C_{W^2}$.  The later equations consistently contain
$C_4+4C_{W^2}$; we therefore treat the intervening factor of two as an
isolated typographical slip.}

The quotient construction complements gauge-invariant characteristic
analyses of four-derivative gravity--matter EFTs
\cite{Reall:2021voz,Davies:2021frz}.  It concerns the order-reduced
low-energy branch.  Recent work on regularized formulations, initial data,
and the weak hyperbolicity of unreduced higher-derivative systems emphasizes
that these notions must not be conflated
\cite{Figueras:2024dta,Gavassino:2026abc,Thaalba:2026abc}.

\section{Established results and limitations}
\label{sec:claims}

We specialize the published $J=2$ vector--scalar anomalous dimension for
opposite helicities to one real scalar and one Abelian vector,
$\bmu\CQ=-55/(96\pi^2)$.  The order-reduced field-redefinition
analysis rederives the on-shell relation $\CQ=C_4+4\CW$ in the normalization
of that residue.  This fixes the contribution of the selected amplitude
coordinate to the relative tensor--photon cone, including its sign and the
normalization chain from the Wilson coefficient to $r_T-r_\gamma$ and to the
local speed slope.  A common running shift changes both absolute cone
parameters equally, so only their difference is claimed as frame invariant.

For local solution jets with nonzero scalar gradient and electromagnetic
field, the
isolated contact generates nonzero mixed blocks in the unreduced principal
symbol.  The explicit Schur analysis shows that its $BD^{-1}C$ term begins at
second order in the contact, while the Maxwell Ward identity removes the
first-order physical off-diagonal projection.  This checks the quotient and
Schur reduction for the selected contact, but it is not a nontrivial
first-order mixed-eigenvector flow.  On the physical EFT quotient, the
standard factorization criterion
$\ker\Amap\subseteq\ker\Xmap$ tests identifiability.  When the declared
domain contains them, the explicit $F^4$ and curvature--photon directions
give stacked rank two on the corresponding backgrounds, so the specified
amplitude coordinate is insufficient to determine the characteristic
response over that domain.  This statement is relative to $\Wflow$; it does
not imply that the missing beta functions fail to exist or that every
witness direction is generated by a particular fixed theory.

The calculation does not provide a term-complete four-derivative
characteristic beta function, a model-specific flow for a complete Horndeski
EFT, or a nontrivial first-order Schur splitting.  It also does not determine
the missing four-photon or $h\gamma\gamma$ beta components.  No claim is made
about a front velocity or ultraviolet causality.

\section{Conclusions}
\label{sec:conclusions}

For the published opposite-helicity vector--scalar residue, the
order-reduced frame map and characteristic projection give
\begin{align}
 \left.\bmu(r_T-r_\gamma)\right|_{J=2}
 &=-\frac{55}{48\pi^2\Mp^4},
 \nonumber\\
 \left.\bmu(c_T-c_\gamma)\right|_{J=2,\Delta=0}
 &=-\frac{55}{96\pi^2}\frac{\Xb}{\Mp^4}.
\end{align}
The first quantity is invariant under a common running shift of all cone
parameters.  The second is its local speed normalization on the baseline
common cone.  Both are scheme- and background-specified contributions, not
threshold-matched observables or absolute lower bounds.

The electromagnetic-background analysis clarifies what this input does not
accomplish.  Although the contact produces nonzero principal mixed blocks
before reduction, $BD^{-1}C$ starts at second order in the contact and the
Maxwell Ward identity annihilates the first-order physical off-diagonal
projection.  Independently, the kernel criterion applied to the declared EFT
and characteristic quotients shows that the single amplitude coordinate
cannot determine the response over a declared domain that includes the
independent $F^4$ and curvature--photon classes: these directions have
vanishing images under $\Amap$ and generate the explicit rank-two matrices
in Eqs.~\eqref{eq:em-kernel-witness} and \eqref{eq:cff-rank}.

The result is therefore both positive and limited.  It supplies a
reproducible contribution to the frame-invariant cone difference and a
concrete analysis of how on-shell information enters characteristic data,
while showing, over the declared domain, that the chosen input is
insufficient to reconstruct a term-complete flow without additional
amplitude data or model information restricting $\Wflow$ to $\Wbeta$.  No
conclusion is drawn here about a complete Horndeski beta function, a front
velocity, or ultraviolet causality.

\begin{acknowledgments}
This research received no specific grant from any funding agency in the
public, commercial, or not-for-profit sectors.
The author declares no competing interests.

The author used OpenAI ChatGPT and Codex (versions available through the
service during July--August 2026; finer backend identifiers were not
consistently exposed) for symbolic cross-checks, literature organization,
and English/LaTeX drafting.  The systems were directed with explicit
normalization conventions, source equations, scheme choices, and claim
boundaries.  No AI output was treated as a scientific source; retained
results were checked against the displayed algebra, cited primary sources,
and accompanying verification script.  The author takes full responsibility
for the derivations, citations, and scientific claims; the AI systems are not
authors.
\end{acknowledgments}

\section*{Data Availability}
No new observational data were generated.  The algebraic verification code,
coefficient-provenance record, and build instructions are supplied as
Supplemental Material with this submission \cite{SupplementalMaterial}.

\appendix

\section{Standard factorization lemma}
\label{app:theorem}

Let $A:E\to D$ and $X:E\to C$ be linear maps.  The following statements are
equivalent:
\begin{enumerate}
\item $X(e)$ depends only on $A(e)$;
\item there exists a linear map $\widetilde X:\operatorname{im}A\to C$ such
that $X=\widetilde X\circ A$;
\item $\ker A\subseteq\ker X$.
\end{enumerate}
The implication $2\Rightarrow3$ is immediate.  For $3\Rightarrow2$, define
$\widetilde X(Ae)=Xe$.  If $Ae=Ae'$, then $e-e'\in\ker A\subseteq\ker X$, so
the definition is independent of the representative.  Finally
$2\Rightarrow1$ is immediate, and $1\Rightarrow3$ follows by comparing $e$
with $e+k$ for $k\in\ker A$.  Applying this result to
$E=\Wflow$, $A=\Amap$, and $X=\Xmap$ proves
Eq.~\eqref{eq:identifiability}.

A scale-independent invertible change of EFT coordinates $U:E\to E$ replaces
the maps by $A\circ U^{-1}$ and $X\circ U^{-1}$ and transports both kernels by
the same isomorphism.  A scale-independent invertible change of physical-root
coordinates acts only in the codomain of $X$ and does not change its kernel.
An explicitly scale-dependent EFT-coordinate or scheme change carries its
own beta-function transformation and defines a correspondingly transformed
inference problem.  This establishes the coordinate invariance claimed in
Sec.~\ref{sec:method}.

\section{Sign and frame check}
\label{app:sign}

If $C(\Lambda)=0$, then for $\mu<\Lambda$,
\begin{equation}
 C(\mu)=\bmu C\ln\frac\mu\Lambda
 =\frac{55}{96\pi^2}\ln\frac\Lambda\mu>0.
\end{equation}
In the tensor frame, $r_\gamma=-2C/\Mp^4<0$ and $r_T=0$.  For
$\bar V^2=-\Xb$,
\begin{equation}
 c_A^2=\frac1{1-r_A\Xb},
\end{equation}
so the photon is slower than the tensor branch in the infrared within the
local EFT.  In the matter frame, $r_\gamma=0$ and $r_T=2C/\Mp^4$.  The
relative result is identical.

\section{Why the characteristic is not a front velocity}
\label{app:front}

The principal symbol of a derivative-truncated local EFT controls
characteristics only inside its frequency window.  A front velocity is the
$\omega\to\infty$ limit of the complete nonlocal form factor and generally
lies outside that window.  No front-causality conclusion follows from the
signs or numbers in this paper \cite{Shore:2007um,Hollowood:2008kq,
deRham:2020zyh}.  The frequency-dependent curved-space QED refractive index
provides an explicit example of this distinction \cite{Hollowood:2008kq}.

\section{Reproducibility scope}
\label{app:reproducibility}

The accompanying script checks the residue arithmetic, normalization chain,
the Schur-order statement in Eq.~\eqref{eq:schur-zero}, both finite rank
tests, and the Jacobs Weyl example.  These checks begin from the displayed
quadratic forms; they do not replace derivation of the Hessians,
field-redefinition map, or Ward identity from the action.  The provenance
record separates imported coefficients from algebra derived in this paper.

OpenAI ChatGPT and Codex (service versions available during July--August
2026; finer backend identifiers were not consistently exposed) assisted with
symbolic cross-checks and with generating and debugging the verification
script.  They were directed using the displayed quadratic forms, explicit
normalization and sign conventions, exact-arithmetic targets, and stated
claim boundaries.  Retained outputs were verified against the analytic
derivations, exact script results, and cited primary sources.  Their broader
role in manuscript preparation is disclosed in the Acknowledgments.

\bibliography{characteristic_cone_identifiability}

\end{document}